\documentclass[aoas,nameyear,dvips]{arximspdf}
\usepackage{algorithm}
\usepackage{graphicx}

\doi{10.1214/10-AOAS380}
\volume{5}
\issue{1}
\pubyear{2011}
\firstpage{124}
\lastpage{149}
\setattribute{copyright}{owner}{\textup{In the Public Domain}}

\makeatletter
\newproclaim{remark}{Remark}[section]
\makeatother

\begin{document}
\begin{frontmatter}

\title{An autoregressive approach to house price~modeling\thanksref{T1}}
\runtitle{An autoregressive approach to house price modeling}

\begin{aug}
\author[A]{\fnms{Chaitra H.} \snm{Nagaraja}\corref{}\ead[label=e1]{chaitra.nagaraja@census.gov}},
\author[B]{\fnms{Lawrence D.} \snm{Brown}\thanksref{t2}\ead[label=e2]{lbrown@wharton.upenn.edu}}
\and
\author[B]{\fnms{Linda H.} \snm{Zhao}\ead[label=e3]{lzhao@wharton.upenn.edu}}
\runauthor{C. H. Nagaraja, L. D. Brown and L. H. Zhao}
\affiliation{US Census Bureau, University of Pennsylvania and University of Pennsylvania}
\thankstext{T1}{Disclaimer: This report is released to inform interested parties of research and to encourage
discussion. The views expressed on statistical issues are those of the authors and not necessarily those of the US Census Bureau.}
\thankstext{t2}{Supported in part by NSF
Grant DMS-07-07033.}
\address[A]{C. H. Nagaraja\\
US Census Bureau\\
Center for Statistical Research\\
and Methodology\\
4600 Silver Hill Rd.\\
Washington, DC 20233 \\
USA\\
\printead{e1}}                        %adresu isvedimo komanda gale!
\address[B]{L. D. Brown\\
L. H. Zhao\\
Statistics Department \\
The Wharton School \\
University of Pennsylvania \\
400 Jon M. Huntsman Hall \\
3730 Walnut St. \\
Philadelphia, Pennsylvania 19104-6302 \\
USA\\
\printead{e2}\\
\phantom{E-mail:\ }\printead*{e3}}
\end{aug}

% HISTORY:
\received{\smonth{4} \syear{2009}}
\revised{\smonth{6} \syear{2010}}

% ABSTRACT
\begin{abstract}
A statistical model for predicting individual house prices and
constructing a house price index is proposed utilizing information
regarding sale price, time of sale and location (ZIP code).  This model
is composed of a fixed time effect and a random ZIP (postal) code
effect combined with an autoregressive component.  The former two
components are applied to all home sales, while the latter is applied
only to homes sold repeatedly. The time effect can be converted into a
house price index. To evaluate the proposed model and the resulting
index, single-family home sales for twenty US metropolitan areas from
July 1985 through September 2004 are analyzed.  The model is shown to
have better predictive abilities than the benchmark S\&P/Case--Shiller
model, which is a repeat sales model, and a conventional mixed effects
model.  Finally, Los Angeles, CA, is used to illustrate a historical
housing market downturn.
\end{abstract}

% KEYWORDS
\begin{keyword}
\kwd{Housing index}
\kwd{time series}
\kwd{repeat sales}.
\end{keyword}

\end{frontmatter}

%%%%%%%%%%%%%%%%%%%%%%%%%%%%%%%%%%%%%%%%%%%%%%%%%%%%%%%%%%%%%%%%%%%%%%%%%
%%%%%%%%%%%%%%%%%%%%%%%%%%%%%%%%%%%%%%%%%%%%%%%%%%%%%%%%%%%%%%%%%%%%%%%%%
%s1 ###
\section{Introduction}
%%%%%%%%%%%%%%%%%%%%%%%%%%%%%%%%%%%%%%%%%%%%%%%%%%%%%%%%%%%%%%%%%%%%%%%%%
%%%%%%%%%%%%%%%%%%%%%%%%%%%%%%%%%%%%%%%%%%%%%%%%%%%%%%%%%%%%%%%%%%%%%%%%%
% first paragraph
Modeling house prices presents a unique set of challenges.  Houses
are distinctive, each has its own set of hedonic characteristics:
number of bedrooms, square footage, location, amenities and so
forth.  Moreover, the price of a house, or the value of the bundle of
characteristics, is observed only when sold.  Sales, however, occur
infrequently.  As a result, during any period of time, out of the
entire population of homes, only a small percentage are actually sold.
From this information, our objective is to develop a
practical model to predict prices from which we can construct a price
index.  Such an index would summarize the housing market
and would be used to monitor changes over time.  Including both objectives allows one to
look at both micro and macro features of a market, from individual houses
to entire markets.  In the following discussion, we propose an
autoregressive model which is a simple, but effective and
interpretable, way to model house prices and construct an index.  We
show that our model outperforms, in a predictive sense, the
benchmark S\&P/Case--Shiller Home Price Index method when applied to
housing data for twenty US metropolitan areas.  We use these results to
evaluate the proposed autoregressive model as well as the resulting house price index.

% introduce the repeat sales model
A common approach for modeling house prices, called repeat sales,
utilizes homes that sell multiple times to track market trends.
Bailey, Muth and Nourse~(\citeyear{BMN1963}) first
proposed this method and Case and Shiller (\citeyear{CS1987}, \citeyear{CS1989}) extended it
to incorporate heteroscedastic errors. In both models, the log price
difference between two successive sales of a home is used to
construct an index using linear regression.  The previous
sale price acts as a surrogate for hedonic information, provided the
home does not change substantially between sales.  There is a large body of work
focused on improving the index estimates produced by the Bailey, et al. approach.
For instance, a modified form of the repeat sales model is used for the Home Price
Index produced by the Office of Federal Housing Enterprise Oversight
(OFHEO).  Gatzlaff and Haurin
(\citeyear{GH1997}) suggest a repeat sales model that corrects for the correlation
between economic conditions and the chance of a sale occurring.
Alternatively, Shiller (\citeyear{Shiller1991}) and Goetzmann and Peng (\citeyear{GP2002})
propose arithmetic
average versions of the repeat sales estimator as an alternative to
the original geometric average estimator. The former work is
used commercially by Standard and Poors
 to produce the S\&P/Case--Shiller Home Price Index.  We will be
 using this index in our analysis as it is
the most well known.

Several criticisms have been made about repeat sales methods.
Theoretically, for a house to be included in a repeat
sales analysis, no changes must have been made to it; however,
in practice, that is almost never the case.  Furthermore, Englund,
Quigley and Redfearn (\citeyear{EQR1999}) and Goetzmann and Speigel (\citeyear{GS1995}) have
commented on the difficulty of detecting such changes without the
availability of additional information about the home.  Goetzmann
and Speigel, however, do propose an
alternate model which corrects for the effect of changes to homes around
the time the house is sold.

Even if homes which have changed are removed from the data set, an index constructed out of
the remaining homes may still not reflect the true index value.
Case and Quigley (\citeyear{CQ1991}) argue that houses age which has a depreciating effect
on their price.  Therefore, as Case, Pollakowski and Wachter (\citeyear{CPW1991}) write, repeat
sales indices produce estimates of time effects confounded with
age effects.  Palmquist~(\citeyear{Palmquist1982}) has suggested
applying an independently computed depreciation factor to account for the impact of age.

In a sample period, out of the entire population of homes, only a small
fraction are actually sold.  A fraction of these sales are repeat sales
homes with no significant changes.  Recall that the remaining sales, those of
the single sales homes, are omitted from the analysis.  If repeat
sales indices are used to describe the housing market as a
whole, one would like the sample of repeat sales homes to have
similar characteristics to all homes.  If not, Case, Pollakowski and Wachter
remark that the indices would be affected by sample selection bias. Englund, Quigley and Redfearn
in a study of Swedish home sales, and Meese and
Wallace (\citeyear{MW1997}), in a study of Oakland and Freemont home sales, both found that
repeat sales homes are indeed different from single sale homes. Both
studies also observed that in addition to being older, repeat sales homes
were smaller and more ``modest'' [Englund, Quigley and Redfearn (\citeyear{EQR1999})].
Therefore, repeat sales indices seem to provide information only about a
very specific type of home and may not apply to the entire housing
market. However, published indices do not seem to be interpreted in
that manner.  Case and Quigley (\citeyear{CQ1991}) propose an alternative hybrid model that combines repeat
sales methodology with hedonic information which makes use of all sales.
While the index constructed with this method represents all home sales,
it requires housing characteristics which may be
difficult to collect on a broad scale.

We feel the repeat sales concept is valuable
although the current models of this type have the issues described above. The
proposed model applies the repeat sales idea in a new way to
address some of the criticisms while still maintaining the
simplicity and reduced data requirements that the original Bailey et al.
method had. While our primary goal is prediction,
we believe the resulting index could be a better general
description of housing sales than traditional repeat sales
methodology.

\setcounter{footnote}{2}
In our method, log prices are modeled as the sum of a time effect (index), a location effect
modeled as a random effect for ZIP (postal) code, and an underlying first-order
autoregressive time series [AR(1)]. This structure offers four advantages.  First, the price index is estimated with all
sales: single and repeat.  Essentially, the index can be thought of as a weighted
sum of price information from single and repeat sales.  The latter
component receives a much higher weight because more useful
information is available for those homes. Second, the
previous sale price becomes less useful the longer it has been since
the last sale.  The AR(1) series includes this feature into
the model more directly than the Case--Shiller method. Third,
metropolitan areas are diverse and neighborhoods may have disparate
trends.  We include ZIP code effects to model these differences in
location.\footnote{ZIP code was readily available in our data; other
geographic variables at roughly this scale might have been even more
useful had they been available.}  Finally, the proposed model is
straightforward to interpret even while including the features
described above.  We believe the model captures trends in the
overall housing market better than existing repeat sales methods and
is a practical alternative.

We apply this model to data on single family home sales from July 1985
through September 2004 for twenty US metropolitan areas.  These data are described in
Section \ref{sec:data}.  The autoregressive model is outlined and
estimation using maximum likelihood is described in
Section \ref{sec:modelfit}; results are discussed in Section \ref{sec:estimate}.
For comparison, two alternative models are fit: a conventional mixed effects model
and the method used in the S\&P/Case--Shiller Home Price Index.  As a quantitative
way to compare the indices, the predictive capacity of the three methods
are assessed in Section \ref{sec:validate}.   In
Section \ref{sec:la} we examine the case of Los Angeles, CA, where the
proposed model does not perform as well.  We end with a general
discussion in Section \ref{sec:discussion}.

%%%%%%%%%%%%%%%%%%%%%%%%%%%%%%%%%%%%%%%%%%%%%%%%%%%%%%%%%%%%%%%%%%%%%%%%%
%%%%%%%%%%%%%%%%%%%%%%%%%%%%%%%%%%%%%%%%%%%%%%%%%%%%%%%%%%%%%%%%%%%%%%%%%
%s2 ###
\section{House price data}\label{sec:data}
%%%%%%%%%%%%%%%%%%%%%%%%%%%%%%%%%%%%%%%%%%%%%%%%%%%%%%%%%%%%%%%%%%%%%%%%%
%%%%%%%%%%%%%%%%%%%%%%%%%%%%%%%%%%%%%%%%%%%%%%%%%%%%%%%%%%%%%%%%%%%%%%%%%

The data are comprised of single family home sales qualifying for
conventional mortgages from the twenty US metropolitan areas listed
in Table \ref{tab:citylist}.  These sales occurred between July 1985
and September 2004.  Not included in these data are homes
with prices too high to be considered for a conventional
mortgage or those sold at subprime rates.  Note, however, that
subprime loans were not prevalent during the time period covered by
our data.  Similar data are used by Fannie Mae, Freddie
Mac,  and to construct the OFHEO Home Price Index.

For
each sale, the following information is available: address
with ZIP code, month and year of sale, and price.  To ensure
adequate data per time period, we divide the sample period into
three month intervals for a total of 77 periods, or quarters. We
make an attempt to remove sales which are not arm's length by omitting homes
sold more than once in a single quarter.  Given the lack of hedonic information, we have
no way of determining whether a house has changed substantially between sales.  Therefore, we do not
filter our data to remove such houses.

%t1 ###
\begin{table}[b]
\caption{Metropolitan areas in the data}\label{tab:citylist}
\begin{tabular*}{\textwidth}{@{\extracolsep{\fill}}llll@{}}
\hline
Ann Arbor, MI  &    Kansas City, MO  & Minneapolis, MN  &  Raleigh, NC       \\
Atlanta, GA    &    Lexington, KY    & Orlando, FL      &  San Francisco, CA \\
Chicago, IL    &    Los Angeles, CA  & Philadelphia, PA &  Seattle, WA       \\
Columbia, SC   &    Madison, WI      & Phoenix, AZ      &  Sioux Falls, SD   \\
Columbus, OH   &    Memphis, TN      & Pittsburgh, PA   &  Stamford, CT      \\
\hline
\end{tabular*}
\end{table}

%t2 ###
\begin{table}[t]\tablewidth=225pt
\caption{Summary counts for a selection of cities}\label{tab:shortsales}
\begin{tabular*}{225pt}{@{\extracolsep{\fill}}lcc@{}}
\hline
      \textbf{Metropolitan area} & \textbf{Sales} & \textbf{Houses} \\
      \hline
      Stamford, CT   & \phantom{0}14,602 & \phantom{0}11,128 \\
      Ann Arbor, MI  & \phantom{0}68,684 & \phantom{0}48,522  \\
      Pittsburgh, PA & 104,544& \phantom{0}73,871  \\
      Los Angeles, CA& 543,071& 395,061 \\
      Chicago, IL    &688,468 &483,581  \\
      \hline
\end{tabular*}
\end{table}

%t3 ###
\begin{table}[b]\tablewidth=270pt
\caption{Sale frequencies for a selection of cities}
\label{tab:shortrepeats}
\begin{tabular*}{270pt}{@{\extracolsep{\fill}}lcccc@{}}\hline
        \multicolumn{1}{@{}c}{\textbf{Metropolitan area}}  & \multicolumn{1}{c}{\textbf{1 sale}}
        & \multicolumn{1}{c}{\textbf{2 sales}}& \multicolumn{1}{c}{\textbf{3 sales}} & \textbf{4$\bolds{+}$ sales} \\
        \hline
      Stamford, CT    &\phantom{00}8,200   & \phantom{00,}2,502 & \phantom{00,}357   &  \phantom{00,}62 \\
      Ann Arbor, MI   &\phantom{0}32,458  &  \phantom{0,}12,662 &  \phantom{0}2,781&  \phantom{0,}621 \\
      Pittsburgh, PA  &\phantom{0}48,618  & \phantom{0,}20,768  & \phantom{0}3,749 &  \phantom{0,}718 \\
      Los Angeles, CA & 272,258           & 100,918            & 18,965           &  2,903\\
      Chicago, IL     &319,340              &130,234           &28,369            & 5,603 \\
      \hline
\end{tabular*}
\end{table}

\begin{figure}[t]

\includegraphics{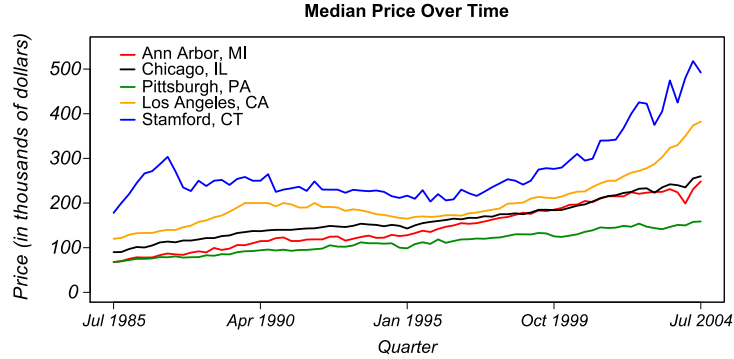}

\caption{Median prices for a selection of cities.}\label{fig:mediantime}
\end{figure}

Table \ref{tab:shortsales} displays the number of sales and unique
houses sold in the sample period for a selection of cities. Complete tables for all summaries in this section
 are provided in Appendix \ref{app:datatable}.  Observe that the total number of sales is always greater
than the number of houses because houses can sell multiple times (repeat sales).  Perhaps more illuminating
is Table \ref{tab:shortrepeats}, where we count the number of times each house is sold.  We see that as the number of sales per
house increases, the number of houses reduces rapidly. Nevertheless,
a significant number of houses sell more than twice.
With a sample period of nearly twenty years, this is not
unusual; however, single sales are the most common despite the long
sample period. The first column of Table \ref{tab:shortrepeats}
shows this clearly.  Moreover, this pattern holds for all cities in our data.
Finally, in Figure \ref{fig:mediantime}, we plot the median price across time for the subset of cities.
This graph illustrates that both the cost of homes and the trends over time vary considerably across cities.

%f1 ###

For  all metropolitan areas in our data, the time of a sale is
fuzzy, as there is often a lag between the day when the price is
agreed upon and the day the sale is recorded (around 20--60 days).
Theoretically, the true value of the house would have changed
between these two points. Therefore, in the strictest sense, the
sale price of the house does not reflect the price at the time when
the sale is recorded. Dividing the year into quarters reduces the
importance of this lag effect.

%%%%%%%%%%%%%%%%%%%%%%%%%%%%%%%%%%%%%%%%%%%%%%%%%%%%%%%%%%%%%%%%%%%%%%%%%
%%%%%%%%%%%%%%%%%%%%%%%%%%%%%%%%%%%%%%%%%%%%%%%%%%%%%%%%%%%%%%%%%%%%%%%%%
%s3 ###
\section{Model}\label{sec:modelfit}
%%%%%%%%%%%%%%%%%%%%%%%%%%%%%%%%%%%%%%%%%%%%%%%%%%%%%%%%%%%%%%%%%%%%%%%%%
%%%%%%%%%%%%%%%%%%%%%%%%%%%%%%%%%%%%%%%%%%%%%%%%%%%%%%%%%%%%%%%%%%%%%%%%%

% Model with technical discussion

The log house price series is modeled as the sum of an index
component, an effect for ZIP code (as an indicator for location),
and an AR(1) time series.  The sale prices of a particular house are treated as a series of sales:
$y_{i,1,z},y_{i,2,z},\ldots,y_{i,j,z},\ldots,$ where $y_{i,j,z}$ is the log sale price of
the $j$th sale of the $i$th house in ZIP code $z$. Note that $y_{i,1,z}$ is
defined as the first sale price in the \textit{sample period}; as a
result, both new homes and old homes sold for the first time in the
sample period are indicated with the same notation.

Let there be $1,\ldots,T$ discrete time periods where house sales occur.  Allow
$t(i,j,z)$ to denote the time period when the $j$th sale of the $i$th
house in ZIP code $z$ occurs and let $\gamma(i,j,z)=t(i,j,z)-t(i,j-1,z)$,
or the gap time between sales.  Finally,
there are a total of $N=\sum_{z=1}^{Z}\sum_{i=1}^{I_z}J_{i}$
observations in the data where there are $Z$ ZIP codes, $I_z$ houses
in each ZIP code and $J_i$ sales for a given house.

The log sale price $y_{i,j,z}$ can now be described as follows:
%
%e1 ###
\begin{eqnarray}\label{eq:ranef}
  y_{i,1,z} & = & \mu + \beta_{t(i,1,z)}+\tau_{z} +
  \varepsilon_{i,1,z}, \qquad j=1, \nonumber\\
  y_{i,j,z} &= & \mu +
  \beta_{t(i,j,z)}+\tau_{z}+ \phi^{\gamma(i,j,z)}\bigl(y_{i,j-1,z}-\mu-\beta_{t(i,j-1,z)}-\tau_{z}\bigr)\\
  &&{}+\varepsilon_{i,j,z}, \qquad j>1,\nonumber
\end{eqnarray}
where:
\begin{enumerate}
\item The parameter $\beta_{t(i,j,z)}$ is the log price index at time $t(i,j,z)$.
Let $\beta_1,\ldots,\beta_T$ denote the log price indices, assumed to be fixed effects.

\item $\phi$ is the autoregressive coefficient and $|\phi|<1$.

\item $\tau_z$ is the random effect for ZIP code $z$.
$\tau_{z}\stackrel{\mathrm{i.i.d.}}{\sim}\mathcal{N}(0, \sigma^{2}_{\tau})$ where $\tau_1,\ldots,\tau_Z$
are the ZIP code random effects which are distributed normally with
mean 0 and variance $\sigma^{2}_{\tau}$ and where i.i.d. denotes
independent and identically distributed.

\item We impose the restriction that $\sum_{t=1}^{T}n_t\beta_t=0$ where
$n_t$ is the number of sales at time $t$. This allows us to interpret $\mu$ as an overall mean.

\item Finally, let
\[
\varepsilon_{i,1,z}\sim\mathcal{N}\biggl(0,
\frac{\sigma^{2}_{\varepsilon}}{1-\phi^2}\biggr),\qquad
\varepsilon_{i,j,z}\sim\mathcal{N}\biggl(0,\frac{\sigma^{2}_{\varepsilon}
(1-\phi^{2\gamma(i,j,z)})}{1-\phi^2}\biggr),
\]
and assume that all $\varepsilon_{i,j,z}$ are independent.
\end{enumerate}

Note that there is only one process for the series $y_{i,1,z},
y_{i,2,z}, \ldots.$  The error variance for the first sale,
$\sigma^{2}_{\varepsilon}/(1-\phi^2)$, is a marginal variance.  For
subsequent sales, because we have information about previous sales, it is
appropriate to use the conditional variance (conditional on the
previous sale),
$\sigma^{2}_{\varepsilon}(1-\phi^{2\gamma(i,j,z)})/(1-\phi^2)$,
instead.  For more details refer to the supplemental article [Nagaraja, Brown and Zhao (\citeyear{NBZ2010})].

The underlying series for each house is given by
$u_{i,j,z}=y_{i,j,z}-\mu-\beta_{t(i,j,z)}-\tau_z$.   We can rewrite
this series as
$u_{i,j,z}=\phi^{\gamma(i,j,z)}u_{i,j-1,z}+\varepsilon_{i,j,z}$
where $\varepsilon_{i,j,z}$ is as given above.  This autoregressive
series is stationary, given a starting observation $u_{i,1,z}$,
because $E[u_{i,j,z}]=0$, a constant, where $E[\cdot]$ is the
expectation function, and the covariance between two points depends
only on the gap\vspace*{1pt} time and not on the actual sale times.  Specifically, $\operatorname{Cov}(u_{i,j,z},
u_{i,j^{\prime},
z})=\sigma_{\varepsilon}^{2}\phi^{(t(i,j^{\prime},z)-t(i,j,z))}/(1-\phi^2)$
if $j<j^{\prime}$.  Therefore, the covariance between a pair of sales
depends \textit{only} on the gap time between sales. Consequently, the
time of sale is uninformative for the underlying series, only the gap
time is required. As a result, the autoregressive series $u_{i,j,z}$ where $i$ and
$z$ are fixed and $j\geq 1$ is a Markov process.

The autoregressive component adds two important features to the model.
Intuitively, the longer the gap time between sales, the less useful
the previous price should become when predicting the next sale
price.  For the model described in (\ref{eq:ranef}),  as the gap
time increases, the autoregressive coefficient decreases by
construction $( \phi^{\gamma(i,j,z)})$, meaning that sales
prices of a home with long gap times are less correlated with each
other.  (See Remark \ref{rem:phi} at the end of this section for
additional discussion on the form of $\phi$.)  Moreover, as the gap time
increases, the variance of the error term increases.  This indicates
that the information contained in the previous sale price is less
useful as the time between sales grows.

To fit the model, we formulate the autoregressive model in (\ref{eq:ranef}) in matrix form:
%
%e2 ###
\begin{equation}\label{eq:matranef}
\mathbf{y} = \mathbf{X}\bolds{\beta}+\mathbf{Z}\bolds{\tau}+\bolds{\varepsilon}^{*},
\end{equation}
where $\mathbf{y}$ is the vector of log prices and $\mathbf{X}$ and
$\mathbf{Z}$ are the design matrices for the fixed effects
$\bolds{\beta}=[\mu\beta_1\ \cdots\ \beta_{T-1}]^{\prime}$
and random effects $\bolds{\tau}$, respectively.  Then, the log
price can be modeled as a mixed effects model with autocorrelated
errors, $\bolds{\varepsilon}^{*}$, and with covariance matrix
$\mathbf{V}$.

We apply a transformation matrix $\mathbf{T}$ to the model in
(\ref{eq:matranef}) to simplify the computations; essentially, this
matrix applies the autoregressive component of the model to both
sides of (\ref{eq:matranef}). It is an $N\times N$ matrix and is
defined as follows. Let $t_{(i,j,z),(i^{\prime}, j^{\prime},
z^{\prime})}$ be the cell corresponding to the $(i,j,z)$th row and
$(i^{\prime}, j^{\prime}, z^{\prime})$th column.  Then,
%e3 ###
\begin{equation}\label{eq:T}
t_{(i,j,z),(i^{\prime}, j^{\prime}, z^{\prime})} =
\cases{
1, &\quad if $i=i^{\prime}$, $j=j^{\prime}$, $z=z^{\prime}$,\cr
-\phi^{\gamma(i,j)}, &\quad if $i=i^{\prime}$, $j=j^{\prime}+1$, $z=z^{\prime}$,\cr
 0,  &\quad otherwise.}
\end{equation}
As a result, $\mathbf{T}\bolds{\varepsilon}^{*}\sim\mathcal{N}
(\mathbf{0},\frac{\sigma^{2}_{\varepsilon}}{1-\phi^2}\operatorname{diag}(\mathbf{r}))$
where $\operatorname{diag}(\mathbf{r})$ is a diagonal matrix of dimension $N$ with the diagonal elements $\mathbf{r}$ being given by
%
%e4 ###
\begin{equation}\label{eq:r}
r_{i,j,z} =
\cases{1, &\quad when $j=1$, \cr
1-\phi^{2\gamma(i,j)}, &\quad when $j>1$.
}
\end{equation}
Using the notation from (\ref{eq:ranef}),
let $\bolds{\varepsilon}=\mathbf{T}\bolds{\varepsilon}^{*}$.
Finally,  we restrict $\sum_{t=1}^{T}n_t\beta_t=0$ where~$n_t$ is the
number of sales at time $t$.  Therefore, $\beta_T = -\frac{1}{n_T}\sum_{t=1}^{T-1}n_t\beta_t$.

The likelihood function for the transformed model is
%
%e5 ###
\begin{eqnarray}\label{eq:locallike}
L(\bolds{\theta};\mathbf{y}) &=&
(2\pi)^{-N/2}|\mathbf{V}|^{-1/2}\nonumber\\ [-8pt]\\ [-8pt]
    &&\times{} \exp\bigl\{-\tfrac{1}{2}\bigl(\mathbf{T}(\mathbf{y}-\mathbf{X}\bolds{\beta})\bigr)^{\prime}\mathbf{V}^{-1}
\bigl(\mathbf{T}(\mathbf{y}-\mathbf{X}\bolds{\beta})\bigr)\bigr\}, \nonumber
\end{eqnarray}
where $\bolds{\theta}=\{\bolds{\beta},\sigma^{2}_{\varepsilon},\sigma^{2}_{\tau},\phi\}$
is the vector of parameters, $N$ is the total number of
observations, $\mathbf{V}$ is the covariance matrix, and
$\mathbf{T}$ is the transformation matrix. We can split $\mathbf{V}$
into a sum of the variance contributions from the time series and
the random effects. Specifically,
%
%e6 ###
\begin{equation}\label{eq:definev}
\mathbf{V}=\frac{\sigma^{2}_{\varepsilon}}{1-\phi^2}\operatorname{diag}(\mathbf{r})+(\mathbf{TZ})\mathbf{D}(\mathbf{TZ})^{\prime},
\end{equation}
where $\mathbf{D}=\sigma^{2}_{\tau}\mathbf{I}_Z$
and $\mathbf{I}_Z$ is an identity matrix with dimension $Z\times Z$.

We use the coordinate ascent algorithm to compute the maximum
likelihood estimates (MLE) of $\bolds{\theta}$ for the model in
(\ref{eq:ranef}).  This iterative procedure maximizes the likelihood
function with respect to each group of parameters while holding all
other parameters constant. The algorithm
terminates when the parameter estimates have converged according to
the specified stopping rule. Bickel and Doksum~(\citeyear{BD2001}) include a proof showing that,
for models in the exponential family,
the estimates computed using the coordinate ascent algorithm converge to the
MLE.  The proposed model, however, is a member of the differentiable
exponential family; therefore, as Brown (\citeyear{Brown1986}) states, the proof does not directly apply.
Nonetheless, we find empirically
that the likelihood function is well behaved, so the MLE appears to
be reached for this case as well. Empirical evidence of convergence can be found in the supplemental
article [Nagaraja, Brown and Zhao (\citeyear{NBZ2010})].

We outline Algorithm \ref{algo1} below.
The equations for updating the parameters and random effects
estimates are given in Appendix \ref{app:update}.

\begin{algorithm}[b]
\caption{Autoregressive (AR) model fitting algorithm.}\label{algo1}
\begin{enumerate}
  \item Set a tolerance level $\bolds{\epsilon}$ (possibly different for each parameter).
  \item Initialize the parameters:
  $\bolds{\theta}^{0}=\{\bolds{\beta}^{0},\sigma^{2,0}_{\varepsilon},\sigma^{2,0}_{\tau},\phi^{0}\}$.
\item For iteration $k$ ($k=0$ when the parameters are initialized),
\begin{enumerate}
    \item[(a)]  Calculate $\bolds{\beta}^{k}$ using (\ref{eq:betalocal}) in Appendix \ref{app:update}
      with $\{\sigma^{2,k-1}_{\varepsilon},\sigma^{2,k-1}_{\tau},\phi^{k-1}\}$.
    \item[(b)]  Compute $\sigma^{2,k}_{\varepsilon}$ by computing the zero of (\ref{eq:sigmaepslocal}) using
      $\{\bolds{\beta}^{k}, \sigma^{2,k-1}_{\tau},\phi^{k-1}\}$.
    \item[(c)]  Compute $\sigma^{2,k}_{\tau}$ by calculating the zero of
    (\ref{eq:sigmataulocal})
    using $\{\bolds{\beta}^{k}, \sigma^{2,k}_{\varepsilon},\phi^{k-1}\}$.
    \item[(d)]  Find the zero of  (\ref{eq:philocal}) to compute $\phi^{k}$ using
    $\{\bolds{\beta}^{k},\sigma^{2,k}_{\varepsilon},\sigma^{2,k}_{\tau}\}$.
\item[(e)]  If $|\bolds{\theta}_{i}^{k-1}-\bolds{\theta}_{i}^{k}|>\bolds{\epsilon}$
    for any $\theta_i\in\bolds{\theta}$, repeat step 3 after replacing
$\bolds{\theta}^{k-1}$ with $\bolds{\theta}^{k}$.
Otherwise, stop (call this iteration $K$).
\end{enumerate}
\item Solve for $\beta_T$ by computing: $\hat{\beta}_T =
-\frac{1}{n_T}\sum_{t=1}^{T-1}n_t\hat{\beta}_{t}^{K}$.
\item Plug in $\{\bolds{\beta}^{K},\sigma^{2,K}_{\varepsilon},
    \sigma^{2,K}_{\tau},\phi^{K} \}$ to
    compute the estimated values for $\bolds{\tau}$ using (\ref{eq:blupblock}).
\end{enumerate}                                                 %%%%%%%%%%%%%%%%%%%%%%%%%%%%
\end{algorithm}

To predict a log price, we substitute the estimated parameters and random effects into (\ref{eq:ranef}):
%
%e7 ###
\begin{equation}
  \hat{y}_{i,j,z} =
  \hat{\mu}+\hat{\beta}_{t(i,j,z)}+\hat{\tau}_z+\hat{\phi}^{\gamma(i,j,z)}\bigl(
  y_{i,j-1,z}-\hat{\mu}-\hat{\beta}_{t(i,j-1,z)}-\hat{\tau}_z
  \bigr).
\end{equation}
We then convert $\hat{y}_{i,j,z}$ to the price scale (denoted as
$\hat{Y}_{i,j,z}$) using
%
%e8 ###
\begin{equation}\label{eq:convertprice}
\hat{Y}_{i,j,z}(\sigma^{2}) =
\exp\biggl\{\hat{y}_{i,j,z}+\frac{\sigma^{2}}{2}\biggr\},
\end{equation}
where $\sigma^2$ denotes the variance of $y_{i,j,z}$. The additional
term $\sigma^2/2$ approximates the difference between
$E[\exp\{X\}]$ and $\exp\{E[X]\}$ where $E[\cdot]$ is the
expectation function.  We must adjust the latter expression to
approximate the conditional mean of the response, $\mathbf{y}$.  We improve the
efficiency of our estimates by using the adjustment stated in Shen, Brown and Zhi (\citeyear{SHZ2006}).
In (\ref{eq:convertprice}), $\sigma^2$ can be
estimated from the mean squared residuals (MSR), where
$\mbox{MSR}=\frac{1}{N}\sum_{i=1}^{N}(y_{i,j,z}-\hat{y}_{i,j,z} )^2$ and $N$ is the total
number of observations used to fit the model. Therefore, the log price estimates,
$\hat{y}_{i,j,z}$, are converted to the price scale by
%
%e9 ###
\begin{equation}\label{eq:eff}
  \hat{Y}_{i,j,z}=\exp\biggl\{\hat{y}_{i,j,z}+\frac{\mbox{MSR}}{2} \biggr\}.
\end{equation}
Goetzmann (\citeyear{Goetzmann1992}) proposes a similar transformation for the index
values computed using a traditional repeat sales method. Calhoun
(\citeyear{Calhoun1996}) suggests applying Goetzmann's adjustment
when using an index value to
predict a particular house price.  For the autoregressive model,
the standard error of the index is sufficiently small that the
efficiency adjustment has a negligible impact on the estimated index.
Therefore, we simply use $\exp\{\hat{\beta}_t \}$ to convert the index to the price scale.  Finally, we
rescale the vector of indices so that the first quarter has an index
value of 1.

\begin{remark}\label{rem:phi}
The autoregressive coefficient form, $\phi^{\gamma(i,j,z)}$,
deserves further explanation.  For each house indexed by $(i,z)$,
let $t_1(i,z)=t(i,1,z)$ denote the time of the initial sale.
Conditioning on the (unobserved) values of the parameters
$\{\mu,\beta_t,\sigma^{2}_{\varepsilon},\sigma^{2}_{\tau}\}$ and
on the values of the random ZIP code effects, $\{\tau_z\}$,
let $\{u_{i,z;t}\dvtx t=t_1(i,z),t_1(i,z)+1,\ldots\}$ be an underlying AR(1)
process.  To be more precise, $u_{i,z;t}$ is a conventional, stationary AR(1) process defined by
%
%e10 ###
\begin{equation}
u_{i,z;t} = \cases{
\varepsilon_{i,1,z}, &\quad if $t=t_1(i,z)$, \cr
\phi u_{i,z;t-1}+\varepsilon_{i,1,z}, &\quad if $t>t_1(i,z)$,
  }
\end{equation}
where if $t=t(i,j,z)$, then $\varepsilon_{i,z;t(i,j,z)}=\varepsilon_{i,j,z}$
and otherwise $\varepsilon_{i,z;t}\stackrel{\mathrm{i.i.d.}}{\sim}\mathcal{N}(0, \frac{\sigma^{2}_{\varepsilon}}{1-\phi^2})$.
Then the observed log sale prices are given\vspace*{1pt} by
$\{y_{i,j,z}\}$ where $u_{i,z;t(i,j,z)}=y_{i,j,z}-(\mu+\beta_{t(i,j,z)}+\tau_z)$.
The values of $u_{i,z;t}$ are to be interpreted as the
\textit{potential sale price adjusted} by $\{\mu,\beta_t,\sigma^{2}_{\varepsilon},\sigma^{2}_{\tau}\}$
of the house indexed by $(i,z)$ \textit{if} the house were to be sold at time $t$.

For housing data like ours, the value of the autoregressive parameter
$\phi$ for this latent process will be near the largest possible value, $\phi=1$.
Consequently, if the underlying process were actually an observed process
from which one wanted to estimate $\phi$, then estimation of $\phi$ could
be a delicate matter.  However, sales generally occur with fairly large gap
times and so the values of $\phi^{\gamma(i,j,z)}$ occurring in the data
will generally not be close to 1.  For that reason, conventional estimation
procedures perform satisfactorily when estimating $\phi$.  We provide
empirical evidence for this in Section \ref{sec:estimate} and in the
supplemental article [Nagaraja, Brown and Zhao~(\citeyear{NBZ2010})].
\end{remark}

%%%%%%%%%%%%%%%%%%%%%%%%%%%%%%%%%%%%%%%%%%%%%%%%%%%%%%%%%%%%%%%%%%%%%%%%%
%%%%%%%%%%%%%%%%%%%%%%%%%%%%%%%%%%%%%%%%%%%%%%%%%%%%%%%%%%%%%%%%%%%%%%%%%
%s4 ###
\section{Estimation results}\label{sec:estimate}
%%%%%%%%%%%%%%%%%%%%%%%%%%%%%%%%%%%%%%%%%%%%%%%%%%%%%%%%%%%%%%%%%%%%%%%%%
%%%%%%%%%%%%%%%%%%%%%%%%%%%%%%%%%%%%%%%%%%%%%%%%%%%%%%%%%%%%%%%%%%%%%%%%%

To fit and validate the autoregressive (AR) model, we divide the
observations for each city into training and test sets.  The test
set contains all final sales for homes that sell three or more
times.  Among homes that sell twice, the second sale is added to the
test set with probability $1/2$.  As a result, the test set for each city contains roughly
15\% of the sales.  The remaining sales
(including single sales) comprise the training set.
Table \ref{tab:traintest} in Appendix \ref{app:datatable} lists
the training and test set sizes for each city.  We fit the model on the training set and examine the estimated
parameters.  The test set will be used in Section \ref{sec:validate} to
validate the AR model against two alternatives.

\begin{figure}[b] % figure 2

\includegraphics{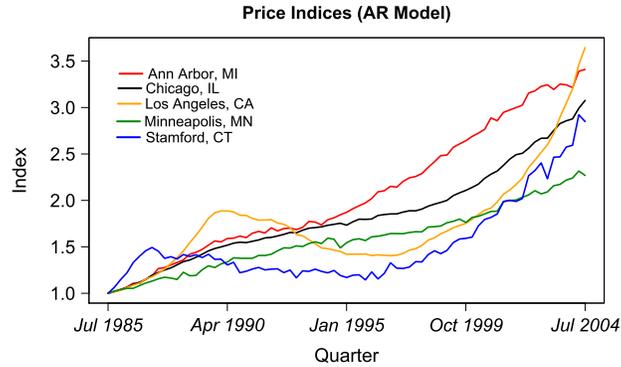}

\caption{The AR index for a selection of cities.}\label{fig:allcityindex}
\end{figure}

In Table \ref{tab:estimateslocal}, the estimates for the overall
mean $\mu$ (on the log scale), the autoregressive parameter $\phi$, the variance of the
error term $\sigma^{2}_{\varepsilon}$, and the variance of the
random effects $\sigma^{2}_{\tau}$ are provided for each
metropolitan area.  As expected, the most expensive cities have the highest
values of $\mu$: Los Angeles, CA, San Francisco, CA, and Stamford,
CT.  In Figure \ref{fig:allcityindex}, the indices for a sample
of the twenty cities are provided.  There are clearly different
trends across cities.
%t4
\begin{table}
\caption{Parameter estimates for the AR model}\label{tab:estimateslocal}
\begin{tabular*}{\textwidth}{@{\extracolsep{\fill}}lcccc@{}}
\hline
    \textbf{Metropolitan area} & $\bolds{\hat{\mu}}$ &$\bolds{\hat{\phi}}$
    & $\bolds{\hat{\sigma}^{2}_{\varepsilon}}$ & $\bolds{\hat{\sigma}^{2}_{\tau}}$ \\
    \hline
    Ann Arbor, MI     & 11.6643 & 0.993247 & 0.001567 & 0.110454 \\
    Atlanta, GA       & 11.6882 & 0.992874 & 0.001651 & 0.070104 \\
    Chicago, IL       & 11.8226 & 0.992000 & 0.001502 & 0.110683 \\
    Columbia, SC      & 11.3843 & 0.997526 & 0.000883 & 0.028062 \\
    Columbus, OH      & 11.5159 & 0.994807 & 0.001264 & 0.090329 \\
    Kansas City, MO   & 11.4884 & 0.993734 & 0.001462 & 0.121954 \\
    Lexington, KY     & 11.6224 & 0.996236 & 0.000968 & 0.048227 \\
    Los Angeles, CA   & 12.1367 & 0.981888 & 0.002174 & 0.111708 \\
    Madison, WI       & 11.7001 & 0.994318 & 0.001120 & 0.023295 \\
    Memphis, TN       & 11.6572 & 0.994594 & 0.001120 & 0.101298 \\
    Minneapolis, MN   & 11.8327 & 0.992008 & 0.001515 & 0.050961 \\
    Orlando, FL       & 11.6055 & 0.993561 & 0.001676 & 0.046727 \\
    Philadelphia, PA  & 11.7106 & 0.991767 & 0.001679 & 0.183495 \\
    Phoenix, AZ       & 11.7022 & 0.992349 & 0.001543 & 0.106971 \\
    Pittsburgh, PA    & 11.3408 & 0.992059 & 0.002546 & 0.103488 \\
    Raleigh, NC       & 11.7447 & 0.993828 & 0.001413 & 0.047029 \\
    San Francisco, CA & 12.4236 & 0.985644 & 0.001788 & 0.056201 \\
    Seattle, WA       & 11.9998 & 0.989923 & 0.001658 & 0.039459 \\
    Sioux Falls, SD   & 11.6025 & 0.995262 & 0.001120 & 0.032719 \\
    Stamford, CT      & 12.5345 & 0.987938 & 0.002294 & 0.093230 \\
    \hline
\end{tabular*}
\end{table}
%f2 ###

The estimates for the AR model parameter $\phi$ are close to one.
This is not surprising as the adjusted log sale prices, $u_{i,j,z}$, for
sale pairs with short gap times are expected to be closer in value than those
with longer gap times.  It may be tempting to assume that since
$\phi$ is so close to 1, the prices form a random walk instead of
an AR(1) time series (see Remark~\ref{rem:phi}).  However, this is clearly not the case. Recall
that $\phi$ enters the model not by itself but as
$\phi^{\gamma(i,j,z)}$ where $\gamma(i,j,z)$ is the gap time.  These
gap times are high enough that the correlation coefficient $\phi^{\gamma(i,j,z)}$ is
considerably lower than 1.  The mean gap time across cities is around 22 quarters. As an example, for Ann Arbor,
MI, $\hat{\phi}^{22}= 0.993247^{22}\approx0.8615$ which is clearly
less than 1.  Therefore, the types of sensitivity often produced as a consequence of
near unit roots do not apply to our autoregressive model.

% Local AR Check plot for Columbus, OH

We have modeled the adjusted log prices,
$u_{i,j,z}=y_{i,j,z}-\beta_{t(i,j,z)}-\tau_{z}$, as a latent AR(1)
time series. Accordingly, for each gap time, $\gamma(i,j,z)=h$, there is an
expected correlation between the sale pairs: $\phi^{h}$.  To
check that the data support the theory, we compare the correlation
between pairs of quarter-adjusted log prices at each gap length
to the correlation predicted by the model.

First, we compute the estimated adjusted log prices
$\hat{u}_{i,j,z}=y_{i,j,z}-\hat{\beta}_{t(i,j,z)}-\hat{\tau}_{z}$ for the training data.
Next, for each gap time $h$, we find all the sale pairs
$(\hat{u}_{i,j-1,z},\hat{u}_{i,j,z})$ with that
particular gap length.  The sample correlation between those sale
pairs produces an estimate of $\phi$ for gap length $h$.  If
we repeat this procedure for each possible gap length, we should obtain a
steady decrease in the correlation as gap time increases.  In
particular, the points should follow the curve~$\phi^h$ if the model
is specified correctly.

\begin{figure} % figure 3

\includegraphics{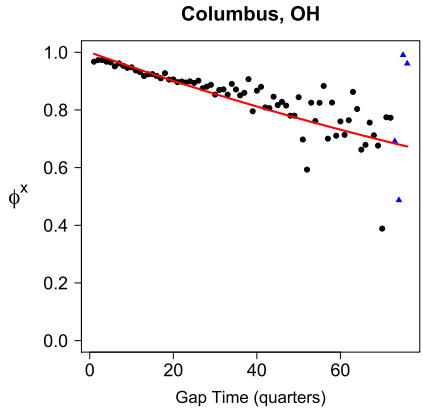}

\caption{Checking the AR(1) assumption for Columbus, OH.}\label{fig:larcolumbus}
\end{figure}

In Figure \ref{fig:larcolumbus}, we plot the correlation of the
adjusted log prices by gap time for Columbus, OH.  Note that the
computed correlations for each gap time were computed with varying
quantities of sale pairs.  Those computed with fewer than twenty sale
pairs are plotted as blue triangles.  We also overlay the predicted
relationship between $\phi$ and gap time.  The inverse relationship
between gap time and correlation seems to hold well and we obtain
similar results for most cities.  One notable exception is Los
Angeles, CA, which we discuss in Section \ref{sec:la}.

%f3 ###

%%%%%%%%%%%%%%%%%%%%%%%%%%%%%%%%%%%%%%%%%%%%%%%%%%%%%%%%%%%%%%%%%%%%%%%%%%
%s5 ###
\section{Model validation}\label{sec:validate}
%%%%%%%%%%%%%%%%%%%%%%%%%%%%%%%%%%%%%%%%%%%%%%%%%%%%%%%%%%%%%%%%%%%%%%%%%%

To show that the proposed AR model produces good predictions, we fit
the model separately to each of the twenty cities and apply the
fitted models to each test set.  For comparison purposes, a mixed effects model
along with the
benchmark S\&P/Case--Shiller model is applied to the data.
The former model is a simple, but
reasonable, alternative to the AR model. Both models are described
below.  In addition to the predictions, we compare the price indices
and training set residuals.

The root mean squared error (RMSE)\footnote{RMSE$=\sqrt{\frac{1}{n}\sum_{k=1}^{n}(Y_k-\hat{Y}_k)^2}$,
where $Y$ is the sale price and $n$ is the test set size.} is used to evaluate predictive performance for each
city in Section \ref{sec:localresult}.  We will see that the AR model
provides the best predictions. In addition, we will show the results from Columbus, OH
as a typical example.

%%%%%%%%%%%%%%%%%%%%%%%%%%%%%%%%%%%%%%%%%%%%%%%
%s5.1 ###
\subsection{Mixed effects model}\label{sec:memtwo}
%%%%%%%%%%%%%%%%%%%%%%%%%%%%%%%%%%%%%%%%%%%%%%%

A mixed effects model provides a very simple, but plausible, approach for
modeling these data.  This model treats the time effect
($\beta_t$) as a fixed effect, and the effects of house ($\alpha_i$)
and ZIP code ($\tau_z$) are modeled as random effects. There is no
time series component to this model. We describe the model as
follows:
%
%e11 ###
\begin{equation}
  y_{i,j,z} = \mu+\alpha_i +\tau_z +\beta_{t(i,j,z)}+\varepsilon_{i,j,z},
\end{equation}
where $\alpha_i\stackrel{\mathrm{i.i.d.}}{\sim}\mathcal{N}(0,
\sigma_{\alpha}^{2})$, $\tau_z\stackrel{\mathrm{i.i.d.}}{\sim}\mathcal{N}(0,
\sigma^{2}_{\tau})$, and
$\varepsilon_{i,j,z}\stackrel{\mathrm{i.i.d.}}{\sim}\mathcal{N}(0,\sigma^{2}_{\varepsilon})$
for houses $i$ from $1,\ldots,I_z$, sales $j$ from $1,\ldots,J_i$,
and ZIP codes $z$ from $1,\ldots,Z$.  As before, $\mu$ is a fixed
parameter and $\beta_{i,j,z}$ is the fixed effect for time.  The
estimates for the parameters $\bolds{\theta}=\{\mu,
\bolds{\beta}, \sigma^{2}_{\varepsilon},\sigma^{2}_{\tau} \}$
are computed using maximum likelihood estimation.

Finally, estimates for the random effects $\bolds{\alpha}$ and
$\bolds{\tau}$ are calculated by iteratively\vadjust{\goodbreak} calculating the
following:
%e13 ###
%e12 ###
\begin{eqnarray}
  \hat{\bolds{\alpha}} &=& \biggl(\frac{\sigma^{2}_{\varepsilon}}{\sigma^{2}_{\alpha}}\mathbf{I}_{I}
                        +\mathbf{W}^{\prime}\mathbf{W}
                        \biggr)^{-1}\mathbf{W}^{\prime}(\mathbf{y}-\mathbf{X}\hat{\bolds{\beta}}-\mathbf{Z}\hat{\bolds{\tau}} ), \\
  \hat{\bolds{\tau}} &=& \biggl(\frac{\sigma^{2}_{\varepsilon}}{\sigma^{2}_{\tau}}\mathbf{I}_{Z}
                        +\mathbf{Z}^{\prime}\mathbf{Z}
                        \biggr)^{-1}\mathbf{Z}^{\prime}(\mathbf{y}-\mathbf{X}\hat{\bolds{\beta}}-\mathbf{W}\hat{\bolds{\alpha}} ),
\end{eqnarray}
where $\mathbf{X}$ and $\mathbf{W}$ are the design matrices for the
fixed and random effects respectively and $\mathbf{y}$ is the
response vector. These expressions are derived using the method of computing
BLUP estimators outlined by Henderson (\citeyear{Henderson1975}).

To predict the log price, $\hat{y}_{i,j,z}$, we substitute the estimated values:
%e14 ###
\begin{equation}
  \hat{y}_{i,j,z} = \hat{\mu}+\hat{\beta}_{t(i,j,z)}+\hat{\alpha}_i+\hat{\tau}_z.
\end{equation}
We use transformation (\ref{eq:eff}) to convert these
predictions back to the price scale.  Finally, we construct a price
index similar to the autoregressive case.  Therefore, as in
Figure \ref{fig:allcityindex}, the values of $\exp\{\hat{\beta}_t \}$
are rescaled so that the price index in the first quarter is 1.

%%%%%%%%%%%%%%%%%%%%%%%%%%%%%%%%%%%%%%%%%%%%%%%
%s5.2 ###
\subsection{S\&P/Case--Shiller model}\label{sec:csdescribe}
%%%%%%%%%%%%%%%%%%%%%%%%%%%%%%%%%%%%%%%%%%%%%%%

The original Case and Shiller (\citeyear{CS1987}, \citeyear{CS1989}) model  is a repeat--sales model which
expands upon the Bailey, Muth and Nourse~(\citeyear{BMN1963}) setting by accounting for heteroscedasticity in
the data due to the gap time between sales.  Borrowing some of
their notation, the framework for their model is
%
%e15 ###
\begin{equation}\label{eq:csmodel}
  y_{i,t} = \beta_t + H_{i,t}+u_{i,t},
\end{equation}
where $y_{i,t}$ is the log price\vspace*{-1pt} of the sale of the $i$th house at
time $t$, $\beta_t$ is the log index at time $t$, and
$u_{i,t}\stackrel{\mathrm{i.i.d.}}{\sim}\mathcal{N}(0,\sigma^{2}_{u})$.
The middle term, $H_{i,t}$, is a Gaussian\vspace*{1pt} random walk which incorporates
the previous log sale price of the house.  Location information, such as ZIP codes, are
not included in this model.  Like the Bailey, Muth and Nourse setup, the Case and Shiller
setting is a model for differences in prices.  Thus, the following
model is fit:
%
%e16 ###
\begin{equation}\label{eq:fittedcsmodel}
  y_{i,t^{\prime}}-y_{i,t} = \beta_{t^{\prime}}-\beta_{t} +\sum_{k=t+1}^{t^{\prime}} v_{i,k} +u_{i,t^{\prime}}-
  u_{i,t},
\end{equation}
where $t^{\prime}>t$.  The random walk steps are normally distributed where
$v_{i,k}\stackrel{\mathrm{i.i.d.}}{\sim}\mathcal{N}(0,\sigma^{2}_{v})$.
Weighted least squares is used to fit the model to account for both
sources of variation.

The S\&P/Case--Shiller procedure
follows in a similar vein but is fit on the price scale instead of the log price scale.
The procedure is similar to the arithmetic index proposed by
Shiller (\citeyear{Shiller1991}) which we will describe next; however, full
details are available in the S\&P/Case--Shiller\tsup{\textregistered}
Home Price Indices: Index Methodology~(\citeyear{SPS2009}) report.  Let there
be $S$ sale pairs, consisting of two consecutive sales of
the same house, and $T$ time periods.  An $S\times (T-1)$
design matrix $\mathbf{X}$, an $S\times(T-1)$ instrumental
variables (IV) matrix $\mathbf{Z}$, and an $S\times 1$
response vector $\mathbf{w}$ are defined next.  Let the
subscripts $s$ and $t$ denote the row and column index
respectively.  Finally, let $Y_{s,t}$ be the sale \textit{price} (not log price) of the house in sale pair $s$ at time $t$.  Therefore, in each sale pair, there will be two prices $Y_{s,t}$ and $Y_{s,t^{\prime}}$ where $t\ne t^{\prime}$.
The matrices \textbf{X}, \textbf{Z} and
vector \textbf{w} where $s$ indicates the
row and $t$ indicates the column are now defined as follows:
\begin{eqnarray*}
X_{s,t} &=&\cases{
-Y_{s,t}, &\quad if first sale of pair $s$ is at time $t$, $t>1$,\cr
Y_{s,t}, &\quad if second sale of pair $s$ is at time $t$,\cr
0, &\quad otherwise,
    }\\
 Z_{s,t}&=&\cases{
 -1, &\quad if first sale of pair $s$ is at time $t$, $t>1$,\cr
  1, &\quad  if second sale of pair $s$ is at time $t$, \cr
  0, & \quad otherwise,
}\\
w_{s} &=&\cases{
Y_{s,t}, &\quad first sale of pair $s$ at time 1,\cr
0, & \quad otherwise.
}
\end{eqnarray*}

The goal is to fit the model
$\mathbf{w}=\mathbf{Xb}+\bolds{\varepsilon}$ where
$\mathbf{b}=(b_1\ \cdots\ b_T)^{\prime}$ is the vector of the reciprocal price
indices.  That is, $B_t=1/b_t$ is the price index at time $t$.  A three-step process is implemented to
fit this model. First, $\mathbf{b}$ is estimated using regression with instrumental variables.
Second, the residuals from this regression are used to compute weights for each observation.
Finally, $\mathbf{b}$ is estimated once more while applying the
weights.  This process, outlined in full in the S\&P/Case--Shiller\tsup{\textregistered}
Home Price Indices: Index Methodology report, is described below:
\begin{enumerate}
\item Estimate $\mathbf{b}$ by running a regression using instrumental
variables: $\hat{\mathbf{b}}=(\mathbf{Z}^{\prime}\mathbf{X})^{-1}\times\mathbf{Z}^{\prime}\mathbf{w}$.
\item Calculate the weights for each observation using the squared
residuals from the first step.  These weights are dependent on the gap
time between sales.  We denote the residual as $\hat{\varepsilon}_i$
which is  an estimate of $u_{i,t^{\prime}}-u_{i,t}+\sum_{k=1}^{t^{\prime}-t} v_{i,k}$.  The expectation of
$\varepsilon_i$ is
$E[u_{i,t^{\prime}}-u_{i,t}+\sum_{k=1}^{t^{\prime}-t} v_{i,k}]=0$ and the variance is
$\operatorname{Var}[u_{i,t^{\prime}}-u_{i,t}+\sum_{k=1}^{t^{\prime}-t}
v_{i,k}]=2\sigma^{2}_{u}+(t^{\prime}-t)\sigma^{2}_{v}$.
To compute\vspace*{1pt} the weights for each observation, the squared
residuals from the first step are regressed against the gap time. That is,
%e17 ###
\begin{equation}\label{eq:secondstage}
\hat{\varepsilon}_{i}^{2} =
\underbrace{\alpha_{0}}_{2\sigma^{2}_{u}}+
\underbrace{\alpha_{1}}_{\sigma^{2}_{v}}(t^{\prime}-t)
+\eta_{i},
\end{equation}
where $E[\eta_{i}]=0$.
The reciprocal of the square root of the fitted values
from the above regression are the weights.  Using their notation, we denote this weight matrix by $\bolds{\Omega}^{-1}$.
\item The final step is to estimate $\mathbf{b}$ again while
incorporating the weights, $\bolds{\Omega}$:
$\hat{\mathbf{b}}=(\mathbf{Z}^{\prime}\bolds{\Omega}^{-1}\mathbf{X})^{-1}
\mathbf{Z}^{\prime}\bolds{\Omega}^{-1}\mathbf{w}$.  The indices are
simply the reciprocals of each element in $\mathbf{b}$ for $t>1$ and,
by construction, $B_1=1$.
\end{enumerate}

Finally, to estimate the prices in the test set, we simply calculate
%e18 ###
\begin{equation}
  \hat{Y}_{i,j} = \frac{\hat{B}_{t(i,j-1)}}{\hat{B}_{t(i,j)}}Y_{i,j-1},
\end{equation}
where $Y_{i,j}$ is the price of the $j$th sale of the $i$th house
and $B_t$ is the price index at time $t$.  We do not apply the
correction proposed by Goetzmann when estimating prices because it
is appropriate only for predictions on the log price scale.
The S\&P/Case--Shiller method is fit on the price scale so no transformation is required.

%%%%%%%%%%%%%%%%%%%%%%%%%%%%%%%%%%%%%%%%%%%%%%%
%s5.3 ###
\subsection{Comparing predictions}\label{sec:localresult}

We fit all three models on the training sets for each city and predict prices for
those homes in the corresponding test set.  The RMSE for the test set observations
is calculated in dollars for each model in order to compare performance across models.
These results are listed in Table \ref{tab:localMSE}.    The model
with the lowest RMSE value for each city is shown in italicized font.  Note that
while the S\&P/Case--Shiller method produces predictions directly on the price scale, the autoregressive
and mixed effects models must be converted back to the price scale using (\ref{eq:eff}).
It is clear
that the AR model performs better than the S\&P/Case--Shiller model
for all of the cities, reducing the RMSE by up to 21\% in some cases;
the AR model produces lower RMSE values when compared to the mixed
effects model as well for nearly all cities, San Francisco, CA, being
the only exception.  Moreover, the AR model performs better
under alternate loss functions as well, which we show in
the supplemental article [Nagaraja, Brown and Zhao
(\citeyear{NBZ2010})].

%t5
%%%%%%%%%%%%%%%%%%%%%%%%%%%%%%%%%%%%%%%%%%%%%%%
\begin{table}[b]\tablewidth=270pt
\caption{Test set RMSE for three models (in dollars)}\label{tab:localMSE}
    \begin{tabular*}{270pt}{@{\extracolsep{\fill}}lccc@{}}
    \hline
    \textbf{Metropolitan area}  & \textbf{AR (local)}
    & \textbf{Mixed effects (local)} & \multicolumn{1}{c@{}}{\textbf{S\&P/C--S}}\\
    \hline
Ann Arbor, MI     & \textit{41,401} & 46,519 & 52,718\\
Atlanta, GA       & \textit{30,914} & 34,912 & 35,482\\
Chicago, IL       & \textit{36,004} & ---    & 42,865\\
Columbia, SC      & \textit{35,881} & 38,375 & 42,301\\
Columbus, OH      & \textit{27,353} & 30,163 & 30,208 \\
Kansas City, MO   & \textit{24,179} & 25,851 & --- \\
Lexington, KY     & \textit{21,132} & 21,555 & 21,731\\
Los Angeles, CA   & \textit{37,438} & ---    & 41,951\\
Madison, WI       & \textit{28,035} & 30,297 & 30,640\\
Memphis, TN       & \textit{24,588} & 25,502 & 25,267\\
Minneapolis, MN   & \textit{31,900} & 34,065 & 34,787\\
Orlando, FL       & \textit{28,449} & 30,438 & 30,158\\
Philadelphia, PA  & \textit{33,246}  & ---    & 35,350\\
Phoenix, AZ       & \textit{28,247} & 29,286 & 29,350\\
Pittsburgh, PA    & \textit{26,406} & 28,630 & 30,135\\
Raleigh, NC       & \textit{25,839} & 27,493 & 26,775\\
San Francisco, CA & 49,927 & \textit{48,217} & 50,249\\
Seattle, WA       & \textit{38,469} & 41,950 & 43,486\\
Sioux Falls, SD   & \textit{20,160} & 21,171 & 21,577\\
Stamford, CT      & \textit{57,722} & 58,616 & 68,132\\
\hline
\end{tabular*}
\end{table}

Note that the RMSE value is missing for Kansas City, MO for the S\&P/Case--Shiller
model.  Some of the observation weights calculated in the second
step of the procedure were negative, halting the estimation process.
This is another drawback to some of the existing repeat sales
procedures.  Calhoun (\citeyear{Calhoun1996}) suggests replacing the sale specific
error $u_{i,t}$ [as given in (\ref{eq:fittedcsmodel})] with a house
specific error $u_{i}$; however, this fundamentally changes the
structure of the error term and, as a result, the fitting process.
Furthermore, it is not implemented in the S\&P/Case--Shiller methodology.
Therefore, we do not apply it to our data.

Three values are also missing in Table \ref{tab:localMSE} for the mixed effect
model results.  For these three cities, the iterative fitting procedure failed
to converge.  We can attribute this to the size of these data and, more importantly,
that the data do not conform well to the mixed effects model structure.

%f4 ###
\begin{figure}

\includegraphics{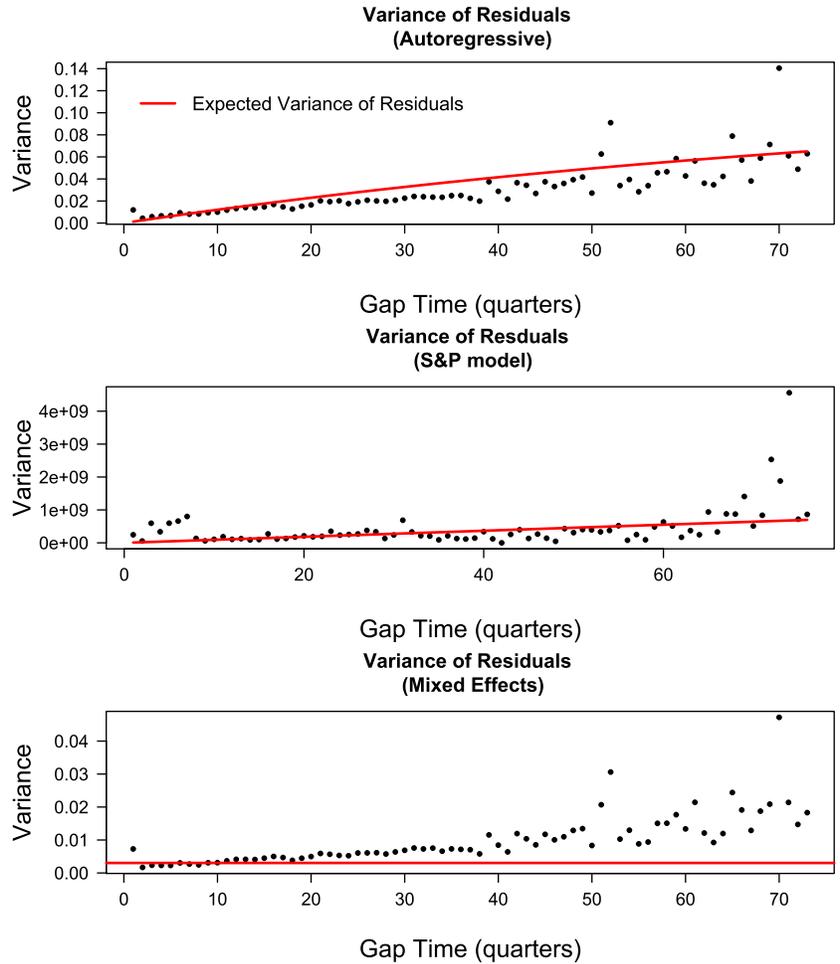}

\caption{Comparing the variance of the residuals for
Columbus, OH.}\label{fig:varplot}
\end{figure}

%f5 ###
\begin{figure} % figure 5

\includegraphics{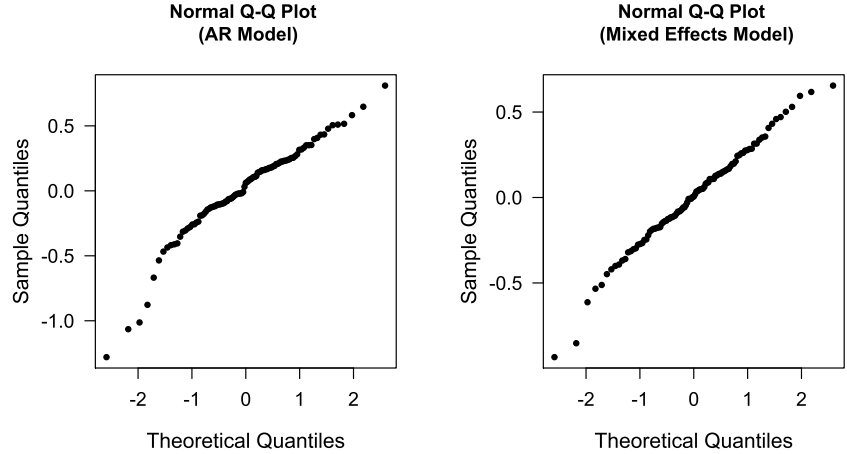}

\caption{Normality of ZIP code effects for Columbus, OH.}\label{fig:columbusraneff}
\end{figure}

Next, we will examine several diagnostic plots to assess whether the
model assumptions are satisfied for each method.  We begin by investigating the variance
of the residuals.  As the gap time increases, we expect a higher
error variance indicating that the previous price becomes less useful over time.
The proposed autoregressive model and the S\&P/Case--Shiller model each
incorporate this feature differently, using an underlying AR(1) time series and a random
walk respectively.  The mixed effects model, however, assumes a
constant variance regardless of gap time.  In Figure \ref{fig:varplot}, for
each model, we plot the variance of the predictions by gap time for the
training set residuals.\footnote{Note that for these three plots,
the term ``residual'' indicates the usual statistical residual values
produced by applying the model and comparing the predictions with
the response vector.  For the AR and mixed effects models, these
residuals are identical to the predictions on the log price scale
discussed in previous sections; however, for the S\&P/C--S model, this is not the case.} The expected variance by gap time values
using the estimated parameters is then overlaid.  The autoregressive
and mixed effects models are fit on the log price scale, whereas the
S\&P/Case--Shiller model is fit on the price scale.  Therefore, the residual plots are graphed on very different scales.

There are two features to note here.  The first is that
heteroscedasticity is clearly present: the variance of the
residuals does in fact increase with gap time.  The second feature
is that while none of the methods perfectly model the
heteroscedastic error, the mixed effects model is undoubtedly the worst.
This pattern holds across all of the cities in the data set.
Both the autoregressive and S\&P/Case--Shiller models seem to have lower than expected variances in Figure~\ref{fig:varplot}.
% Variance of the estimates by gap time

% Normality of Random Effects
        For both the AR and mixed effects models, the random
        effects for ZIP codes are assumed to be normally
        distributed. As a diagnostic procedure, we construct
        the normal quantile plots of the ZIP code effects.  The results are shown in
        Figure~\ref{fig:columbusraneff}. Columbus, OH has a total of 103 ZIP codes, or random effects.
        We find the normality assumption appears to be reasonably satisfied for
        the mixed effects model but less so for the autoregressive model.  Note,
        however, that each random effect is estimated using a different number of sales.
        This interferes with the routine interpretation of these
        plots. In particular, the outliers in both plots
        correspond to ZIP codes containing ten or fewer sales.
        Across all metropolitan areas, the normality assumption seems to
        be well satisfied in some cases and not so well in others, but with no
        clear pattern we could discern as to the type of analysis, size of the data or geographic region.
The supplemental article contains results of the Shapiro--Wilk test for normality [Nagaraja, Brown and Zhao (\citeyear{NBZ2010})].

%f6 ###
\begin{figure} % figure 6

\includegraphics{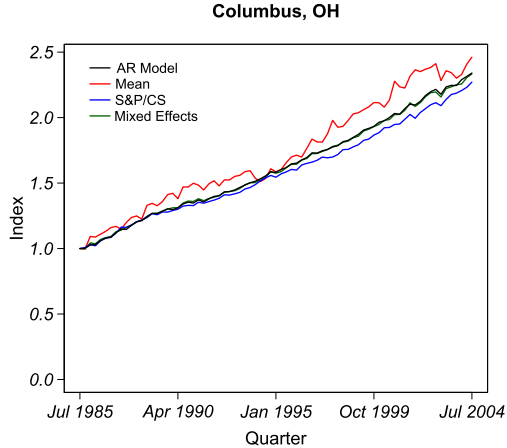}

\caption{House price indices for Columbus, OH.}\label{fig:localindex}
\end{figure}

% Comparing Indices
In Figure \ref{fig:localindex}, we plot four indices for Columbus, OH: the AR index,
the mixed effects index, the S\&P/Case--Shiller index,
and the mean price index.  The mean index is simply
the average sale price at each quarter rescaled so that the first index
value is 1.  From the plot, we see that the autoregressive index is
generally between the S\&P/Case--Shiller index and the mean
index at each point in time.  The mean index treats all sales as single sales. That is,
information about repeat sales is not included; in fact, no
information about house prices is shared across quarters. The
S\&P/Case--Shiller index, on the other hand, only includes repeat
sales houses.  The autoregressive model, because it includes both single
sales and repeat sales, is a mixture of the two perspectives.
Essentially, the index constructed from the proposed model is a
measure of the average house price placing more weight to those homes
which have sold more than once.

%%%%%%%%%%%%%%%%%%%%%%%%%%%%%%%%%%%%%%%%%%%%%%%%%%%%%%%%%%%%%%%%%%%%%%%%%
%%%%%%%%%%%%%%%%%%%%%%%%%%%%%%%%%%%%%%%%%%%%%%%%%%%%%%%%%%%%%%%%%%%%%%%%%
%s6 ###
\section{The case of Los Angeles, CA}\label{sec:la}
%%%%%%%%%%%%%%%%%%%%%%%%%%%%%%%%%%%%%%%%%%%%%%%%%%%%%%%%%%%%%%%%%%%%%%%%%
%%%%%%%%%%%%%%%%%%%%%%%%%%%%%%%%%%%%%%%%%%%%%%%%%%%%%%%%%%%%%%%%%%%%%%%%%

\begin{figure}% figure 7

\includegraphics{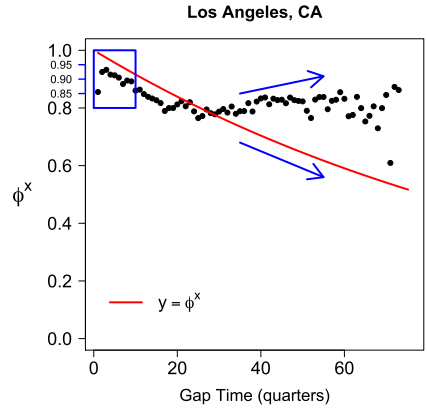}

\caption{Problems with the assumptions.}\label{fig:lawrong}
\end{figure}

Even though the autoregressive model has a lower RMSE than the
S\&P/Case--Shiller model for Los Angeles, CA, it does not seem to fit the data
well.  If we examine Figure \ref{fig:lawrong}, a plot of the
correlation against gap time, we immediately see two significant
issues when what is expected (line) is compared with what the data indicate (dots).  First, the value of $\phi$
is not as close to 1 as expected.  Second, the pattern of decay, $\phi^{\gamma(i,j,z)}$, also
does not follow the presumed pattern.  We will focus on Los Angeles, CA, and discuss these
two issues for the remainder of this section.

%f7 ###

\begin{figure} % figure 8

\includegraphics{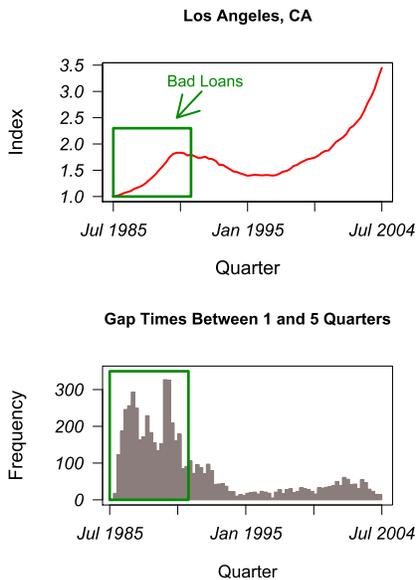}

\caption{Examining the housing downturn.}\label{fig:indexla}
\end{figure}

We expect $\phi$ to be close to 1; however, for Los
Angeles, CA, this does not seem to be the case.  In fact, according
to the data, for short gap times, the correlation between sale pairs
seems to be  far lower than one.  To investigate this feature, we examine
sale pairs with gap times
between 1 and 5 quarters more closely.  In Figure~\ref{fig:indexla},
we construct a histogram of the quarters where the second sale
occurred for this subset of sale pairs.  We pair this histogram with a
plot of the price index for Los Angeles, CA.  Most of these sales
occurred during the late 1980s and early 1990s.  This corresponds to
the same period when Sing and Furlong (\citeyear{SF1989}) found that lenders were offering people mortgages where
the monthly payment was \textit{greater} than 33\% of their monthly
income.  The threshold of 33\% is set to help ensure
that people will be able to afford their mortgage.  Those persons
with mortgages that exceed this percentage tend to have a higher
probability of defaulting on their payments.

%f8 ###

Bates (\citeyear{Bates1989}) found that a number of banks including the Bank
of California and Wells Fargo were highly exposed to these
risky investments, especially in the wake of the housing
downturn during the early 1990s.  If a short gap
time is an indication that a foreclosure took place, this would
explain why these sale pair prices are not highly correlated.
We did observe, however, that other cities also experienced
periods of decline, such as Stamford, CT (see Figure \ref{fig:allcityindex}),
but did not have anomalous autoregressive patterns  like those in Figure \ref{fig:lawrong} for Los Angeles,
CA.

Even if this were not the case, the autoregressive model may not be
performing well simply because there was a downturn in the housing
market. Most of the cities in our data cover periods where the
indices are increasing--the model may be performing well only
because of this feature.  In the case of Los Angeles, CA,
if we examine the period between January 1990 and December 1996 on
Figure \ref{fig:indexla}, the housing index was decreasing.  However, if we calculate the RMSE
of test set sales for this period only, we find that
the autoregressive model still performs better than the
S\&P/Case--Shiller method.  The RMSE values are \$32,039
and \$41,841, respectively.  Therefore, the autoregressive model
seems to perform better in a period of decline as well as in times of
increase.

The second irregularity evident in Figure \ref{fig:lawrong} is that the AR(1)
process does not decay at the same rate as the model predicts.  In
1978 California voters, as a protest against rising property taxes,
passed Proposition 13 which limited how fast property tax
assessments could increase per year.  Galles and Sexton (\citeyear{GS1998}) argue
that Proposition 13 encouraged people to retain homes especially if they
have owned their home for a long time. It is possible that this feature of
Figure \ref{fig:lawrong} is a long term effect of Proposition 13.  On
the other hand, it could be that California home owners tend to
renovate their homes more frequently than others, reducing the decay
in prices over time.  However, we have no way of verifying either of
these explanations given our data.

%%%%%%%%%%%%%%%%%%%%%%%%%%%%%%%%%%%%%%%%%%%%%%%%%%%%%%%%%%%%%%%%%%%%%%%%%
%%%%%%%%%%%%%%%%%%%%%%%%%%%%%%%%%%%%%%%%%%%%%%%%%%%%%%%%%%%%%%%%%%%%%%%%%
%s7 ###
\section{Discussion}\label{sec:discussion}
%%%%%%%%%%%%%%%%%%%%%%%%%%%%%%%%%%%%%%%%%%%%%%%%%%%%%%%%%%%%%%%%%%%%%%%%%
%%%%%%%%%%%%%%%%%%%%%%%%%%%%%%%%%%%%%%%%%%%%%%%%%%%%%%%%%%%%%%%%%%%%%%%%%

Two key tasks when analyzing house prices are predicting
sale prices of individual homes and constructing price
indices which measure general housing trends.  Using extensive
data from twenty metropolitan areas, we have compared our
predictive method to two other methods, including the S\&P/Case--Shiller
Home Price Index.  We find that on average the predictions using our
method are more accurate in all but one of the twenty metropolitan areas
examined.

Data such as ours often do not contain reliable hedonic information
on individual homes, if at all.  Therefore, harnessing the information
contained in a previous sale is critical.  Repeat sales indices
attempt to do exactly that.  Some methods have also incorporated
ad hoc adjustments to take account of the gap time between the
repeat sales of a home.  In contrast, our model involves an
underlying AR(1) time series which automatically adjusts for
the time gap between sales.  It also uses the home's ZIP code
as an additional indicator of its hedonic value.  This indicator has
some predictive value, although its value is quite weak by
comparison with the price in a previous sale if one has been recorded.

The index constructed from our statistical model can be viewed
as a weighted average of estimates from single and repeat sales
homes, with the repeat sales prices having a substantially
higher weight.  As noted, the time series feature of the model
guarantees that this weight for repeat sales prices slowly
decreases in a natural fashion as the gap time between sales increases.

Our results do not provide definitive evidence as to the value of
our index when comparing with other currently available indices
as a general economic indicator.  Indeed, such a determination
should involve a study of the economic uses of such indicators
as well as an examination of their formulaic construction and
their use for prediction of individual sale prices.  We have
not undertaken such a study, and so can offer only a few comments
about the possible comparative values of our index.

As we have discussed, we feel it may be an advantage that our
index involves all home sales in the data (subject to the naturally
occurring weighting described above), rather than only repeat sales.
Repeat sales homes are only a small, selected fraction of all home sales.
Studies have shown that repeat sales homes may have different
characteristics than single sale homes.  In particular, they are
evidently older on average, and this could be expected to have an
effect on their sale price.  Since our measure brings all home sales
into consideration, albeit in a gently weighted manner, and since it
provides improved prediction on average, it may produce a preferable index.

Another advantage of our model is that it remains easy to interpret
at both the micro and macro levels, in spite of including several
features inherent in the data.  Future work seems desirable to
understand anomalous features such as those we have discussed in
the Los Angeles, CA, area.  Such research may allow us to construct a
more flexible model to accommodate such cases.  For example, it could
involve the inclusion of economic indicators which may affect house
prices such as interest rates and tax rates and measures of general
economic status such as the unemployment rate.

%%%%%%%%%%%%%%%%%%%%%%%%%%%%%%%%%%%%%
%%%%%%%%%%% APPENDICES %%%%%%%%%%%%%%
%%%%%%%%%%%%%%%%%%%%%%%%%%%%%%%%%%%%%

\begin{appendix}

%%%%%%%%%%%%%%%%%%%%%%%%%%%%%%%%%%%%%%%%%%%%%%%%%%%%%%%%%%%%%%%%%%%%%%%%%
%%%%%%%%%%%%%%%%%%%%%%%%%%%%%%%%%%%%%%%%%%%%%%%%%%%%%%%%%%%%%%%%%%%%%%%%%
%s8 ###
\section{Data Summary}\label{app:datatable}
%%%%%%%%%%%%%%%%%%%%%%%%%%%%%%%%%%%%%%%%%%%%%%%%%%%%%%%%%%%%%%%%%%%%%%%%%
%%%%%%%%%%%%%%%%%%%%%%%%%%%%%%%%%%%%%%%%%%%%%%%%%%%%%%%%%%%%%%%%%%%%%%%%%

% MAKE NUMBER OF ZIP CODES INTO A SEPRATE TABLE
%t6 ###
\begin{table}[H]
\vspace*{-10pt}
\caption{Summary counts}\label{tab:datasize}
\begin{tabular*}{\textwidth}{@{\extracolsep{\fill}}lcccccc@{}}
\hline
&&& \multicolumn{4}{c@{}}{\textbf{No. houses per sale count}} \\ [-7pt]
&&& \multicolumn{4}{c@{}}{\hrulefill}\\
\multicolumn{1}{@{}l}{\textbf{City}}&\multicolumn{1}{c}{\textbf{No. sales}} &  \multicolumn{1}{c}{\textbf{No. houses}}
&  \multicolumn{1}{c}{\textbf{1}}  & \multicolumn{1}{c}{\textbf{2}} & \multicolumn{1}{c}{\textbf{3}}  &
\multicolumn{1}{c@{}}{\textbf{4$\bolds{+}$}}\\
 \hline
 Ann Arbor, MI     &    \phantom{0}68,684&  \phantom{0}48,522    &  \phantom{0}32,458  &  \phantom{0}12,662&   \phantom{0}2,781 &    \phantom{0,}621  \\
 Atlanta, GA       &              376,082&            260,703    &            166,646  &  \phantom{0}76,046&             15,163 &              2836 \\
 Chicago, IL       &              688,468&            483,581    &            319,340  &            130,234&             28,369 &             5,603 \\
 Columbia, SC      &    \phantom{00}7,034&  \phantom{00}4,321    &  \phantom{00}2,303  &  \phantom{00}1,470&   \phantom{00,}431 &    \phantom{0,}117 \\
 Columbus, OH      &              162,716&            109,388    &  \phantom{0}67,926  &  \phantom{0}31,739&   \phantom{0}7,892 &             1,831 \\
 Kansas City, MO   &              123,441&  \phantom{0}90,504    &  \phantom{0}62,489  &  \phantom{0}23,706&   \phantom{0}3,773 &    \phantom{0,}534 \\
 Lexington, KY     &    \phantom{0}38,534&  \phantom{0}26,630    &  \phantom{0}16,891  &  \phantom{00}7,901&   \phantom{0}1,555 &    \phantom{0,}282 \\
 Los Angeles, CA   &              543,071&            395,061    &            272,258  &            100,918&             18,965 &             2,903 \\
 Madison, WI       &    \phantom{0}50,589&  \phantom{0}35,635    &  \phantom{0}23,685  &  \phantom{00}9,439&   \phantom{0}2,086 &    \phantom{0,}425 \\
 Memphis, TN       &    \phantom{0}55,370&  \phantom{0}37,352    &  \phantom{0}23,033  &  \phantom{0}11,319&   \phantom{0}2,412 &    \phantom{0,}587 \\
 Minneapolis, MN   &              330,162&            240,270    &            166,811  &  \phantom{0}59,468&             11,856 &             2,127 \\
 Orlando, FL       &              104,853&  \phantom{0}72,976    &  \phantom{0}45,966  &  \phantom{0}22,759&   \phantom{0}3,706 &    \phantom{0,}543 \\
 Philadelphia, PA  &              402,935&            280,272    &            179,107  &  \phantom{0}82,681&             15,878 &             2,606 \\
 Phoenix, AZ       &              180,745&            129,993    &  \phantom{0}87,249  &  \phantom{0}35,910&   \phantom{0}5,855 &    \phantom{0,}968 \\
 Pittsburgh, PA    &              104,544&  \phantom{0}73,871    &  \phantom{0}48,618  &  \phantom{0}20,768&   \phantom{0}3,749 &    \phantom{0,}718 \\
 Raleigh, NC       &              100,180&  \phantom{0}68,306    &  \phantom{0}42,545  &  \phantom{0}20,632&   \phantom{0}4,306 &    \phantom{0,}818 \\
 San Francisco, CA &    \phantom{0}73,598&  \phantom{0}59,416    &  \phantom{0}46,959  &  \phantom{0}10,895&   \phantom{0}1,413 &    \phantom{0,}149 \\
 Seattle, WA       &              253,227&            182,770    &            124,672  &  \phantom{0}47,406&   \phantom{0}9,198 &             1,494 \\
 Sioux Falls, SD   &    \phantom{0}12,439&  \phantom{00}8,974    &  \phantom{00}6,117  &  \phantom{00}2,353&   \phantom{00,}419 &   \phantom{00,}85 \\
 Stamford, CT      &    \phantom{0}14,602&  \phantom{0}11,128    &  \phantom{00}8,200  &  \phantom{00}2,502&   \phantom{00,}357 &   \phantom{00,}62 \\
 \hline
\end{tabular*}
\vspace*{-25pt}
\end{table}
%t7 ###
\begin{table}[H]\tablewidth=170pt
\caption{Number of ZIP codes by city}\label{tab:ZIP}
\begin{tabular*}{170pt}{@{\extracolsep{\fill}}lc@{}}\hline
\textbf{City}              & \textbf{No. ZIP codes} \\ \hline
 Ann Arbor, MI     & \phantom{0}57    \\
 Atlanta, GA       & 184   \\
 Chicago, IL       & 317   \\
 Columbia, SC      & \phantom{0}12    \\
 Columbus, OH      & 103    \\
 Kansas City, MO   & 179   \\
 Lexington, KY     & \phantom{0}31    \\
 Los Angeles, CA   & 280   \\
 Madison, WI       & \phantom{0}40    \\
 Memphis, TN       & \phantom{0}64    \\
  Minneapolis, MN   & 214   \\
   Orlando, FL       & \phantom{0}96    \\
    Philadelphia, PA  & 330   \\
     Phoenix, AZ       & 130   \\
    \hline
\end{tabular*}
\end{table}
\setcounter{table}{6}
\begin{table}\tablewidth=170pt
\caption{(Continued.)}\label{tab:ZIP}
\begin{tabular*}{170pt}{@{\extracolsep{\fill}}lc@{}}\hline
\textbf{City}              & \textbf{No. ZIP codes} \\ \hline
 Pittsburgh, PA    & 257   \\
 Raleigh, NC       & \phantom{0}82    \\
 San Francisco, CA & \phantom{0}70    \\
 Seattle, WA       & 110   \\
 Sioux Falls, SD   & \phantom{0}30    \\
 Stamford, CT      & \phantom{0}23    \\ \hline
\end{tabular*}
\end{table}

%t8 ###
\begin{table}
\caption{Training and test set sizes}\label{tab:traintest}
\begin{tabular*}{\textwidth}{@{\extracolsep{\fill}}lccccc@{}}
\hline
     &\multicolumn{3}{c}{\textbf{Autoregressive model}}&
     \multicolumn{2}{c@{}}{\textbf{S\&P/Case--Shiller model}}\\ [-7pt]
    &\multicolumn{3}{c}{\hrulefill}&\multicolumn{2}{c@{}}{\hrulefill}\\
    \textbf{City} & \textbf{Training} & \textbf{Test} & \textbf{No. houses} & \textbf{Training pairs}
    & \textbf{No. houses}\\ \hline
    Ann Arbor, MI     &   \phantom{0}58,953 & \phantom{0}9,731  & \phantom{0}48,522 & \phantom{0}10,431 & \phantom{0}9,735   \\
    Atlanta, GA       &             319,925 &           56,127  &           260,703 & \phantom{0}59,222 &           55,911  \\
    Chicago, IL       &             589,289 &           99,179  &           483,581 &           105,708 &          99,069  \\
    Columbia, SC      &    \phantom{0}5,747 & \phantom{0}1,287  & \phantom{00}4,321 & \phantom{00}1,426 & \phantom{0}1,279   \\
    Columbus, OH      &             136,989 &           25,727  &           109,388 & \phantom{0}27,601 &           25,458 \\
    Kansas City, MO   &             107,209 &           16,232  & \phantom{0}90,504 & \phantom{0}16,705 &           16,092  \\
    Lexington, KY     &   \phantom{0}32,705 & \phantom{0}5,829  & \phantom{0}26,630 & \phantom{00}6,075 & \phantom{0}5,748   \\
    Los Angeles, CA   &             470,721 &           72,350  &           395,061 & \phantom{0}75,660 &          72,338  \\
    Madison, WI       &   \phantom{0}43,349 & \phantom{0}7,240  & \phantom{0}35,635 & \phantom{00}7,714 & \phantom{0}7,221   \\
    Memphis, TN       &   \phantom{0}46,724 & \phantom{0}8,646  & \phantom{0}37,352 & \phantom{00}9,372 & \phantom{0}8,673   \\
    Minneapolis, MN   &             286,476 &           43,686  &           240,270 & \phantom{0}46,206 &  43,764  \\
    Orlando, FL       &   \phantom{0}89,123 &           15,730  & \phantom{0}72,976 & \phantom{0}16,147 &           15,531  \\
    Philadelphia, PA  &             343,354 &           59,581  &           280,272 & \phantom{0}63,082 &           60,068 \\
    Phoenix, AZ       &             155,823 &           24,922  &           129,993 & \phantom{0}25,830 &         24,656  \\
    Pittsburgh, PA    &   \phantom{0}89,762 &           14,782  & \phantom{0}73,871 & \phantom{0}15,891 &           14,956  \\
    Raleigh, NC       &   \phantom{0}84,678 &           15,502  & \phantom{0}68,306 & \phantom{0}16,372 &           15,388  \\
    San Francisco, CA &   \phantom{0}66,527 & \phantom{0}7,071  & \phantom{0}59,416 & \phantom{00}7,111 & \phantom{0}6,948   \\
    Seattle, WA       &             218,741 &           34,486  &           182,770 & \phantom{0}35,971 &        34,304  \\
    Sioux Falls, SD   &   \phantom{0}10,755 & \phantom{0}1,684  & \phantom{00}8,974 & \phantom{00}1,781 & \phantom{0}1,677   \\
    Stamford, CT      &   \phantom{0}12,902 & \phantom{0}1,700  & \phantom{0}11,128 & \phantom{00}1,774 & \phantom{0}1,654   \\ \hline
\end{tabular*}
\end{table}

%%%%%%%%%%%%%%%%%%%%%%%%%%%%%%%%%%%%%%%%%%%%%%%%%%%%%%%%%%%%%%%%%%%%%%%%%
%%%%%%%%%%%%%%%%%%%%%%%%%%%%%%%%%%%%%%%%%%%%%%%%%%%%%%%%%%%%%%%%%%%%%%%%%%
%s9 ###
\section{Updating Equations}\label{app:update}
%%%%%%%%%%%%%%%%%%%%%%%%%%%%%%%%%%%%%%%%%%%%%%%%%%%%%%%%%%%%%%%%%%%%%%%%%%
%%%%%%%%%%%%%%%%%%%%%%%%%%%%%%%%%%%%%%%%%%%%%%%%%%%%%%%%%%%%%%%%%%%%%%%%%

In this section we provide the updating equations for estimating
the parameters $\bolds{\theta}=\{
\bolds{\beta},\sigma^{2}_{\varepsilon},\sigma^{2}_{\tau},\phi\}$
in the autoregressive model (see Section \ref{sec:modelfit}). Observe
that the covariance matrix $\mathbf{V}$ is an $N\times N$ matrix
where $N$ is the sample size. Given the size of our data, it
is simpler computationally to exploit the block diagonal structure
of $\mathbf{V}$.  Each block, denoted by $\mathbf{V}_{z,z}$,
corresponds to observations in ZIP code $z$. Computations are carried out
on the ZIP code level and the updating equations provided below
reflect this.  For instance, $\mathbf{y}_z$ and $\mathbf{T}_z$ are
the elements of the log price vector and transformation matrix respectively for
observations in ZIP code~$z$.

To start, an explicit expression for $\bolds{\beta}$ can be formulated:
%
%e19 ###
\begin{equation}\label{eq:betalocal}
    \hat{\bolds{\beta}} =
    \Biggl(\sum_{z=1}^{Z}(\mathbf{T}_z\mathbf{X}_z)^{\prime}\mathbf{V}_{z,z}^{-1}\mathbf{T}_z\mathbf{X}_z\Biggr)^{-1}
                \sum_{z=1}^{Z}(\mathbf{T}_z\mathbf{X}_z
    )^{\prime}\mathbf{V}_{z,z}^{-1}\mathbf{T}_z\mathbf{y}_z.
\end{equation}
Estimates must be computed
numerically for the remaining parameters.  As all of these are one-dimensional parameters,
methods such as the Newton--Raphson algorithm are highly suitable. We
first define
$\mathbf{w}_{z}=\mathbf{y}_z-\mathbf{X}_z\bolds{\beta}$ for
clarity. To update~$\sigma^{2}_{\varepsilon}$, compute the zero of
%
%e20 ###
\begin{equation}\label{eq:sigmaepslocal}
  0 =
  -\sum_{z=1}^{Z}\operatorname{tr}(\mathbf{V}_{z,z}^{-1}\operatorname{diag}(\mathbf{r}_z)) +
  \sum_{z=1}^{Z}(\mathbf{T}_z\mathbf{w}_z)^{\prime}
  \mathbf{V}_{z,z}^{-1}\operatorname{diag}(\mathbf{r}_z)\mathbf{V}_{z,z}^{-1}(\mathbf{T}_z\mathbf{w}_z),
\end{equation}
where $\operatorname{tr}(\cdot)$ is the trace of a matrix
and $\operatorname{diag}(\mathbf{r})$ is as defined in (\ref{eq:r}).
Similarly, to update~$\sigma^{2}_{\tau}$, find the zero of
 %
%e21 ###
\begin{eqnarray}\label{eq:sigmataulocal}
  0 &=& \sum_{z=1}^{Z}\operatorname{tr}(\mathbf{V}_{z,z}^{-1}(\mathbf{T}_z\mathbf{1}_{n_z})(\mathbf{T}_z\mathbf{1}_{n_z})^{\prime})
  \nonumber \\ [-8pt]\\ [-8pt]
  &&{} +\sum_{z=1}^{Z}(\mathbf{T}_z\mathbf{w}_z)^{\prime}\mathbf{V}_{z,z}^{-1}
  (\mathbf{T}_z\mathbf{1}_{n_z})(\mathbf{T}_z\mathbf{1}_{n_z})^{\prime}\mathbf{V}_{z,z}^{-1}(\mathbf{T}_z\mathbf{w}_z),\nonumber
\end{eqnarray}
where $n_z$ denotes the number of observations in ZIP code $z$ and $\mathbf{1}_k$ is a $(k \times 1)$ vector of ones.

Finally, to update the autoregressive parameter $\phi$, we must
calculate the zero of the function below:
%e22 ###
\begin{eqnarray}\label{eq:philocal}
    0 &=& -\sum_{z=1}^{Z}
        \operatorname{tr}\biggl\{\mathbf{V}_{z,z}^{-1}\biggl(\sigma^{2}_{\tau}
        \biggl(\frac{\partial (\mathbf{T}_z\mathbf{1}_{n_z})}{\partial\phi} \biggr)
        (\mathbf{T}_z\mathbf{1}_{n_z})^{\prime}\nonumber\\
        &&\hphantom{-\sum_{z=1}^{Z}
        \operatorname{tr}\biggl\{\mathbf{V}_{z,z}^{-1}\biggl(} +\sigma^{2}_{\tau}
        (\mathbf{T}_z\mathbf{1}_{n_z})\biggl(\frac{\partial
        (\mathbf{T}_z\mathbf{1}_{n_z})}{\partial\phi}
        \biggr)^{\prime}\nonumber\\
        &&\hphantom{-\sum_{z=1}^{Z}
        \operatorname{tr}\biggl\{\mathbf{V}_{z,z}^{-1}\biggl(}+ \frac{2\phi\sigma^{2}_{\varepsilon}}{(1-\phi^2)^2}
        \operatorname{diag}(\mathbf{r}_z)+\frac{\sigma^{2}_{\varepsilon}}
        {1-\phi^2}\frac{\partial
        \operatorname{diag}(\mathbf{r}_z)}{\partial\phi}\biggr)\biggr\}\nonumber\\
        &&{} -\sum_{z=1}^{Z} \biggl(\frac{\partial\mathbf{T}_z}{\partial\phi}\mathbf{w}_z
    \biggr)^{\prime}\mathbf{V}_{z,z}^{-1}(\mathbf{T}_z\mathbf{w}_z) -\sum_{z=1}^{Z}(\mathbf{T}_z\mathbf{w}_z)^{\prime}\mathbf{V}_{z,z}^{-1}
        \biggl(\frac{\partial\mathbf{T}_z}{\partial\phi}\mathbf{w}_z\biggr) \nonumber\\
        &&{}+\sum_{z=1}^{Z}\biggl[(\mathbf{T}_z\mathbf{w}_z)^{\prime}\mathbf{V}_{z,z}^{-1}\biggl[\sigma^{2}_{\tau}\biggl(\frac{\partial
(\mathbf{T}_z\mathbf{1}_{n_z})}{\partial\phi}
\biggr)(\mathbf{T}_z\mathbf{1}_{n_z})^{\prime}\\
&&{}+\sigma^{2}_{\tau}
(\mathbf{T}_z\mathbf{1}_{n_z})\biggl(\frac{\partial (\mathbf{T}_z\mathbf{1}_{n_z})}{\partial\phi}
\biggr)^{\prime}+ \frac{2\phi\sigma^{2}_{\varepsilon}}{(1-\phi^2)^2}\operatorname{diag}(\mathbf{r}_z)\nonumber\\
&&\hspace*{121.5pt}{}+\frac{\sigma^{2}_{\varepsilon}}{1-\phi^2}\frac{\partial \operatorname{diag}(\mathbf{r}_z)}{\partial\phi}\biggr]\mathbf{V}_{z,z}^{-1}
    (\mathbf{T}_z\mathbf{w}_z)\biggr].\nonumber
\end{eqnarray}

After the estimates converge, we must estimate the
random effects.  We use Henderson's procedure to derive the
Best Linear Unbiased Predictors (BLUP) for each ZIP code.
His method assumes that the parameters in the covariance
matrix, $\mathbf{V}$, are known; however, we use the estimated
values.  The formula is
%e23 ###
\begin{eqnarray}\label{eq:blupblock}
  \hat{\tau}_z &=& \biggl[\frac{2\hat{\sigma}^{2}_{\varepsilon}}{\hat{\sigma}^{2}_{\tau}}+
                        (1-\hat{\phi}^2)(\hat{\mathbf{T}}_z\mathbf{1}_z)^{\prime}\operatorname{diag}^{-1}
                        (\mathbf{r}_z)(\hat{\mathbf{T}}_z\mathbf{1}_z)\biggr]^{-1}\nonumber
                        \\[-8pt]\\ [-8pt]
                &&\times{}\bigl( (1-\hat{\phi}^2)(\hat{\mathbf{T}}_z\mathbf{1}_z)^{\prime}\operatorname{diag}^{-1}
                        (\mathbf{r}_z)(\hat{\mathbf{T}}_z\hat{\mathbf{w}}_z
                        )\bigr),\nonumber
\end{eqnarray}
where $\operatorname{diag}^{-1}(\hat{\mathbf{r}})$ is the inverse of the estimated diagonal matrix
$\operatorname{diag}(\mathbf{r})$.
\end{appendix}

\section*{Acknowledgments}
The authors would like to thank the referees for their thorough and helpful comments.

% AOS,AOAS: If there are supplements please fill:
\begin{supplement}[id=suppA]
 %\sname{}
  \stitle{Supplement to ``An autoregressive approach to house price modeling''}
  \slink[doi]{10.1214/10-AOAS380SUPP}
  \slink[url]{http://lib.stat.cmu.edu/aoas/380/supplement.pdf}
  \sdatatype{.pdf}
  \sdescription{This supplement contains extra analysis on a
  variety of topics related to the paper from examining the
  convergence of the coordinate ascent algorithm, or applying
  alternate loss functions, to studying the impact of each feature included in the autoregressive (AR) model.}
\end{supplement}

\printaddresses

\end{document}